\documentclass[letter]{aa} 

\usepackage{graphicx}
\usepackage{txfonts}
\usepackage{graphicx}
\usepackage{amsmath}
\usepackage[version=4]{mhchem}
\usepackage{siunitx}
\usepackage{pdflscape}
\usepackage{longtable,tabularx}
\usepackage{lscape}
\usepackage{float}
\usepackage{multirow}
\usepackage{multicol}
\usepackage{url}
\usepackage{comment}
\usepackage{xcolor}
\usepackage{adjustbox}

\begin{document}

\title{Observations of the new meteor shower from comet 46P/Wirtanen}

\titlerunning{46P meteor shower observations}

\author{
D. Vida \inst{1,2}\fnmsep\thanks{\email{dvida@uwo.ca}} \and
J. M. Scott \inst{3} \and
A. Egal \inst{4,1,5} \and
J. Vaubaillon \inst{4} \and
Q.-Z. Ye \inst{6,7}  \and
D. Rollinson \inst{8} \and
M. Sato \inst{9} \and
D. E. Moser \inst{10}
}

\institute{    
Department of Physics and Astronomy, University of Western Ontario, London, Ontario, N6A 3K7, Canada
\and
Institute for Earth and Space Exploration, University of Western Ontario, London, Ontario N6A 5B8, Canada
\and
Department of Geology, University of Otago, Dunedin, New Zealand
\and
Planétarium de Montréal, Espace pour la Vie, 4801 av. Pierre-de Coubertin, Montréal, Québec, Canada
\and
IMCCE, Observatoire de Paris - PSL, 77 av. Denfert-Rochereau 75014 Paris, France
\and
Department of Astronomy, University of Maryland, College Park, MD 20742, USA
\and
Center for Space Physics, Boston University, Boston, MA 02215, USA
\and
Perth Observatory Volunteer Group, Bickley, Western Australia
\and
National Astronomical Observatory of Japan, Tokyo, Japan
\and
NASA Meteoroid Environment Office, Marshall Space Flight Center, Huntsville, AL 35812, USA
}

\date{Received January 29, 2024; accepted February 6, 2024}

 
\abstract
   {A new meteor shower $\lambda$-Sculptorids produced by the comet 46P/Wirtanen was forecast for December 12, 2023. The predicted activity was highly uncertain, but generally considered to be low. Observations in Australia, New Zealand, and Oceania were solicited to help constrain the size distribution of meteoroids in the shower.}
   {This work aims to characterize the new meteor shower, by comparing the observed and predicted radiants and orbits, and to provide a calibration for future predictions.}
   {Global Meteor Network video cameras were used to observe the meteor shower. Multi-station observations were used to compute trajectories and orbits, while single-station observations were used to measure the flux profile.}
   {A total of 23 $\lambda$-Sculptorid orbits have been measured. The shower peaked at a zenithal hourly rate (ZHR) of $0.65^{+0.24}_{-0.20}$ meteors per hour at $\lambda_{\odot} = 259.988^{\circ} \pm 0.042^{\circ}$. Due to the low in-atmosphere speed of 15~km s$^{-1}$, the mean mass of observed meteoroids was 0.5~g ($\sim10$~mm diameter), an order of magnitude higher than predicted. The dynamical simulations of the meteoroid stream can only produce such large meteoroids arriving at Earth in 2023 with correct radiants when a very low meteoroid density of $\sim 100$~kg~m$^{-3}$ is assumed. However, this assumption cannot reproduce the activity profile. It may be reproduced by considering higher density meteoroids in a larger ecliptic plane-crossing time window ($\Delta T$ = 20 days) and trails ejected prior to 1908, but then the observed radiant structure is not reproduced.}
   {}
   \keywords{meteoroids --
                Comets: general --
                method: data analysis
               }

   \maketitle
%

\section{Introduction}\label{sec:intro}

Before launch delays, the comet 46P/Wirtanen was initially selected as the primary target for the Rosetta mission due to its favourable orbit and interesting characteristics \citep{Rickman1998}. At the time, the comet was interesting as it had one of the largest nongravitational accelerations of any comet that had undergone a recent orbital perturbation (i.e., close approaches to Jupiter in 1972 and 1984), indicating a high fraction of actively outgassing area \citep{schulz1999coma, groussin2003activity}. The comet had a very close approach to Earth in 2018, coming within 0.077~AU. \citet{Moulane2023} performed observations using TRAPPIST telescopes and found that the comet is ``hyperactive," with 40\% of its surface active and having the activity parameter $Af\rho(0)$ of 250~cm. In comparison, typical Jupiter-family comets have active areas between 5-10 \% \citep{ahearn1995ensemble}. In a study relevant to this work, \citet{coulson2020james} found the possible presence of mm-sized dust in the 46P's coma, meaning that they could be detected optically upon entering the atmosphere.

The comet's orbit has a very low minimum orbit intersection distance (MOID) to Earth of only 0.071~AU, which made it a target of investigation as a potential meteor shower parent body \citep{YeJenniskens2022}. \citet{MaslovMuzyko2017} predicted possible low-activity outbursts in 2017 and 2019 with a zenithal hourly rate (ZHR) not exceeding 5-10 meteors per hour. No positive confirmations of the outbursts have been made in those years. \citet{vaubaillon2023new} performed a separate set of simulations using five different methods that mostly differ in terms of the conditions of meteoroid ejection from the comet and in the modeling of non-gravitational forces during orbital integration. The dynamical models by the co-authors are referred to in this paper by their initials: AE \citep{Egal2019}, JV \citep{Vaubaillon2005a, Vaubaillon2005b}, QY \citep{Ye2016}, MS \citep{Sato2005}, and DM \citep{Moser2004EMP, Moser2008EMP}). All models roughly agreed and found no significant encounters of ejected particles with the Earth in the past, except in 2007 and 2018 with the 1974 trail. A search on Canadian Meteor Orbit Radar (CMOR) data, which spans over two decades \citep{Brown2010} was performed for these years, as well as the years predicted by \citet{MaslovMuzyko2017}; however, no conclusive detections have been found. However, due to the slow speed of meteoroids, which severely impacts the ionization efficiency and the meteor shower geocentric radiants being under the horizon, only a fraction of the largest meteoroids could even be theoretically detectable due to zenithal attraction \citep{gural2001fully}.

For 2023, all five models of \citet{vaubaillon2023new} predict an encounter with the 1974 trail ranging in time between December 12 08:23 UT and 20:06 UT. In addition, the AE model predicts encounters with trails ejected between 1900 and 1945 a bit later, between December 12 at 17:05 UT and December 13 at 06:26 UT. Due to a lack of calibration on observations, only the upper limit on the ZHR has been provided, estimated not to exceed ten meteors per hour on average. The predicted distribution of geocentric radiants on the sky was roughly diagonal and spanned about 15$^{\circ}$, from approximately R.A. = 10$^{\circ}$, Decl. = -45$^{\circ}$ to R.A. = 6$^{\circ}$, Decl. = -33$^{\circ}$. The large extent of the radiant locations was due to the uncertainty in encounter parameters and included all meteoroids within 0.05 AU from Earth. However, meteoroids that were predicted to come the closest, within 0.001 AU, were all concentrated in a tight 1$^{\circ}$ radius around R.A. = 7.75$^{\circ}$, Decl. = -39$^{\circ}$. The radiant at the nominal peak time on December 12 at 10:15 UT was predicted to be in the zenith just off the southwestern tip of Australia, located at the edge of the Earth's terminator. The best locations for optical observations were thus in Australia, New Zealand, and the rest of the Southwest Pacific. The meteor shower was initially referred to as the "Wirtanenids," but \citet{vaubaillon2023new} suggested a new name for the shower, $\lambda$-Sculptorids, which has been accepted by the International Astronomical Union (IAU) for the name of the shower and this term is used herein to refer to the shower. The provisional designation for the shower by the IAU Meteor Data Center is M2023-Y1.

In this work, we report the first conclusive observations of meteors produced by dust ejected from comet 46P/Wirtanen. In Section \ref{sec:methods}, we describe the hardware and methods used by the Global Meteor Network. In Section \ref{sec:results}, we present the results of observations and the comparison to predictions.

\section{Methods} \label{sec:methods}

The Global Meteor Network (GMN\footnote{GMN website: \url{https://globalmeteornetwork.org/} (accessed Jan 8, 2024)}) is a professional-amateur consortium that currently operates over 1000 video meteor cameras globally \citep{vida2021global}. The network has over 200 cameras in the Southwest Pacific region, mostly operated by Fireballs Aotearoa\footnote{Fireballs Aotearoa: \url{https://fireballs.nz/} (accessed Dec 25, 2023)} in New Zealand, Perth Observatory in Western Australia\footnote{Perth Observatory: \url{https://www.perthobservatory.com.au/} (accessed Dec 25, 2023)}, and a number of independent citizen scientists. An average GMN camera system consists of a commercial off-the-shelf (COTS) video camera with a Sony IMX291 sensor, coupled to a 3.6~mm f/0.95 lens and a Raspberry Pi single-board computer for data capture and processing. The optical setup achieves similar performance to human vision, with a field of view (FoV) of $88^{\circ} \times 48^{\circ}$, a frame rate of 25 frames per second and a stellar limiting magnitude in average sky conditions of $+6.0$ \citep{vida2021global}. In addition, a pair of six-camera stations in Western Australia (Pemberton and Toodyay) were operated with 6~mm lenses with a FOV of $53^{\circ} \times 30^{\circ}$ and a stellar limiting magnitude of $+6.5$.

Individual cameras report to a central server that performs meteor trajectory correlation and meteor shower flux computation. Multi-station trajectories and orbits are computed using the method described in \citet{vida2020estimating}, which provides realistic uncertainties of orbital parameters by using Monte Carlo error estimation. Meteor shower flux is computed from single-station observations using the method described in  \citet{vida2022computing}. The total observed atmospheric area at the height of the meteor shower is estimated by applying bias corrections, which take into account the theoretical radiant elevation, meteor angular velocity, and measured camera sensitivity and vignetting. An algorithm is employed to determine periods of clear skies by matching observed stars to their computed positions and numbers assuming modeled observational biases. The method produces a meteor activity profile in units of meteoroids per 1000 square kilometres at a given mass limit, together with a profile of observation efficiency, which informs the completeness of the results (e.g., a peak might be missing in the activity profile if all cameras were clouded out at that time). In practice, for any new meteor shower, meteor trajectories are computed first and the radiant location, radiant drift, radiant dispersion, meteor velocity, average meteor height, and meteoroid mass distribution index are manually measured. These data are then put in the flux algorithm.

\section{Observation results and comparison to simulations} \label{sec:results}

During the time of the predicted $\lambda$-Sculptorid activity, most of New Zealand was under cloud cover, with only a handful of cameras of Fireballs Aotearoa having clear skies the whole night. The weather in Australia was excellent, with most cameras observing uninterrupted the whole night. Qualitatively speaking, the observed activity was very low, with each camera observing on the order of one $\lambda$-Sculptorid meteor during the whole night. 

\subsection{Trajectories and orbits}

A total of 23 $\lambda$-Sculptorid meteoroid trajectories and orbits were computed, the details of which are given in Table \ref{tab:trajectories}. All were observed in either New Zealand or Australia, except for one each from South Korea and Bulgaria where the radiant was under the horizon but the trajectories were bent by Earth's gravity enabled by the extremely low speed of the shower (i.e., zenithal attraction). The observed median geocentric radiant is $\alpha_g = 6.89^{\circ}$, $\delta_g = -38.89^{\circ}$ in equatorial and $\lambda_g - \lambda_{\odot} = 88.10^{\circ}$, $\beta_g = -38.19^{\circ}$ in Sun-centered ecliptic coordinates. The radiant dispersion, measured as the median angular separation from the median radiant \citep{moorhead2021meteor}, is $0.96^{\circ}$. The median radiant measurement error is $0.52^{\circ}$, meaning that the true physical radiant dispersion has been resolved. The median observed geocentric velocity is $10.00 \pm 0.25$~km~s$^{-1}$.

\begin{figure}[!h]
    \centering
    \includegraphics[width=.5\textwidth]{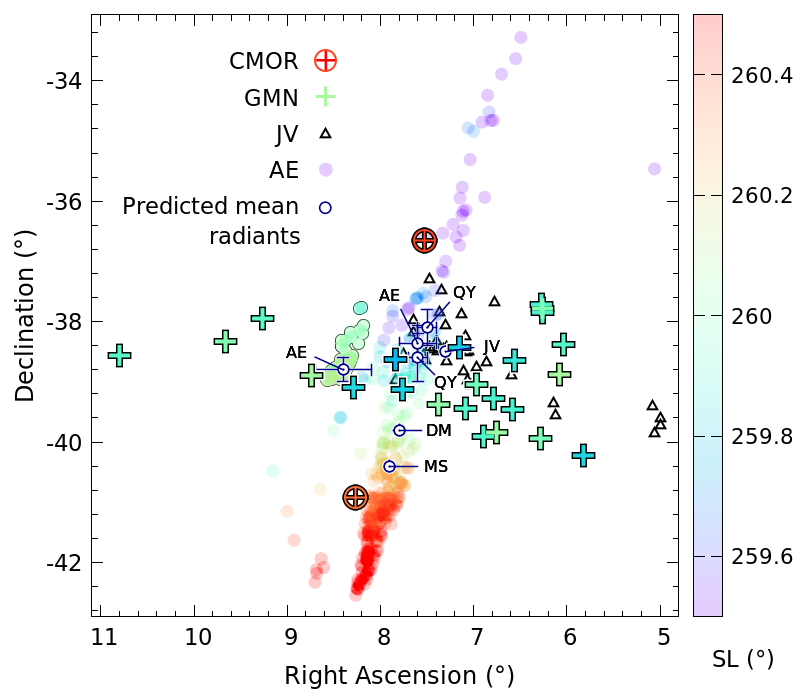}
    \caption{Comparison between the geocentric radiants measured by GMN (crosses) and CMOR (circled crosses) and the various theoretical radiants proposed by \cite{vaubaillon2023new}. The models are marked by the initials of individual co-authors. The coloured circles are radiants predicted by AE, based on the \citet{Egal2019} model, with mm-sized particles having a dark edge (to the left of the smaller particles). Triangles are model radiants predicted by JV using the \citet{Vaubaillon2005a} model. The radiants observed by the GMN are shown as crosses. The error bars are not plotted for better readability, but the average radiant measurement error is $\sim0.5^{\circ}$.}
    \label{fig:radiants_comparison}
\end{figure}

Figure \ref{fig:radiants_comparison} shows the comparison between model radiants and the observed GMN and CMOR radiants. The observed GMN radiant locations generally match well with the models. The mean observed geocentric declination matches almost exactly, with some models predicting it to be at the observed $-38.8^{\circ}$, while the rest are all between $-38.0^{\circ}$ and $-40.4^{\circ}$ for the 1974 trail. However, there is a $\sim 1^{\circ}$ shift in right ascension from the mean model radiant locations. In terms of the spread, the AE model radiants extend about $8^{\circ}$ along the declination, while the JV radiants extend $\sim2^{\circ}$ along the right ascension, matching the GMN video observations best. All models predicted a mean encounter geocentric velocity in the range between 9.8 and 10.3~km~s$^{-1}$, which perfectly matches the observed value of $10.00 \pm 0.25$~km~s$^{-1}$.

In addition to GMN, CMOR observed a total of four $\lambda$-Sculptorid meteors, despite being located in Canada. Two had large measurement uncertainties due to the extremely low speed. The other two have been observed well and presented in the figure. The improved atmospheric deceleration correction for CMOR \citep{froncisz2020possible} enabled accurate measurement of the pre-atmospheric speed of these slow meteors. The electron line densities for the two meteors are around $10^{14}$~e~m$^{-1}$, translating to a peak meteor magnitude of about +5.5 and 3--5~mm diameter particles \citep{weryk2013simultaneous}. To confirm the validity of the detection, a period of 20 days before the shower was investigated for additional detections, and not a single radiant was observed by CMOR within a $20^{\circ}$ radius from the shower during that time. The two radiants (Fig. \ref{fig:radiants_comparison}) were located exactly on the elongated radiant spread modeled by AE, matching that model best.

For reference, the mean orbital parameters using GMN data have been computed using the method of \citet{jopek2006calculation} and are given in Table \ref{tab:mean_orbit} where they are compared to the mean of simulated orbits (JV), and the orbit of the comet 46P/Wirtanen at epochs in 2018 (latest orbit) and the time of meteoroid ejection in 1974. Compared to the comet, both the observed and simulated meteoroids have a smaller orbit, with the perihelion just inside the Earth's orbit. The mean observed and simulated orbits compare well and are well within the scatter in the orbital parameters for both (see Table \ref{tab:trajectories}).

\begin{table*}[]
    \centering
    \caption{Mean orbital elements of the observed $\lambda$-Sculptorids compared to the mean orbital elements of the simulated particles encountering the Earth in 2023 (JV model) and the comet 46P/Wirtanen\protect\footnotemark.}
    \begin{tabular}{l r r r r l}
        \hline
        Parameter & $\lambda$-Sculptorids & Simulated & 46P/Wirtanen & 46P/Wirtanen & \\
        \hline
        Epoch     & J2000       & J2000      & 2458465.5    & 2442240.5 & \\
        Date      & 2023-12-12  & 2023-12-12 & 2018-12-13   & 1974-07-12 & \\
        a         & 2.891       & 3.029      &   3.092662   & 3.255261 & AU \\
        q         & 0.984566    & 0.980229   &   1.05535538 & 1.255701 & AU\\
        e         & 0.65946     & 0.676399   &   0.6587550  & 0.614255 & \\
        i         &   9.200     & 8.364      &  11.747548   & 12.2672  & deg\\
        $\Omega$  &  79.925     & 79.432     &  82.157634   & 84.2087  & deg\\
        $\omega$  & 359.915     & 0.284      & 356.341072   & 351.8580 & deg\\
        \hline
    \end{tabular}
    \label{tab:mean_orbit}
\end{table*}
\footnotetext{JPL Horizons: \url{https://ssd.jpl.nasa.gov/tools/sbdb_lookup.html#/?sstr=46P} (accessed Dec 25, 2023)}

\subsection{Physical properties}

The shower meteors exhibited median begin and end heights of $90.1 \pm 1.2$~km and $78.2 \pm 2.9$~km, respectively. The median meteor peak height is $84.2 \pm 2.5$~km, translating into an F parameter of $0.54 \pm 0.13$. The observed absolute peak magnitudes ranged from the faintest meteor at $+2.40$ to the brightest at $-0.45$. The photometric masses were computed on integrated light curves using a fixed luminous efficiency of $\tau = 0.7\%$ \citep{vida2018modelling}. The observed meteoroids had masses between 0.15 -- 4.12~g, with a median mass of 0.47~g. Using a bulk density of 500~kg~m$^{-3}$ \citep{Ceplecha1998}, these masses correspond to diameters between 8 - 25~mm, and the median value of 12~mm.

We computed the \citet{ceplecha1958composition} $K_B$ parameter to characterize the physical composition. \citet{cordonnier2024not} calibrated the parameter on GMN data and found that a value of $-0.1$ needs to be subtracted from the original definition, a correction that we also applied. The shower had a mean $K_B$ of $6.972 \pm 0.096$, putting the meteors into group C1, defined as regular cometary material from short-period comets \citep{ceplecha1988earth}. This result is consistent with the comet's origin and orbit, making the meteoroids similar to the Taurids, for instance. A similar value of $K_B$ has also been observed for the 2022 outburst of the $\tau$-Herculids \citep{egal2023modeling}. That outburst was caused by similarly-aged meteoroids, ejected in 1995. Both comets were on a similar orbit with a $q \approx 1$ and the measured $K_B$ for the $\tau$-Herculids is $6.9 \pm 0.2$ \citep{koten2023tau}. This is in contrast to the Draconids, which are also young \citep{Egal2019}, and have a parent body on a similar orbit, but are much more fragile, classified as group D \citep{koten2007optical}. This indicates the physical structural strength of meteoroids is not only a function of age, but is inherent to the specificity of the material present in the parent comet.

\subsection{Activity profile}

Due to the small number of observed meteors, the population index has been difficult to measure. Using the method of \citet{vida2020new}, we found a mass index of $s = 2.25^{+1.5}_{-0.52}$ and a population index of $r = 3.23^{+3.03}_{-0.85}$. The errors are large due to the small sample size, making choosing a single nominal value difficult. We erred on the side of caution, choosing values of $s = 2.2$ and $r = 3.0$ which have been traditionally considered in the literature as values appropriate for showers rich in small particles \citep{rendtel2004population}, as indicated from observations of 46P \citep{Kareta2023, vaubaillon2023new}. The chosen value is well within the uncertainty.

Figure \ref{fig:gmn_flux} in the appendix shows the observed flux profile of the $\lambda$-Sculptorids and the data is given in Table \ref{tab:flux}. The shower peaked at solar longitude $\lambda_{\odot} = 259.988^{\circ} \pm 0.042^{\circ}$ (2023-12-12 15:04 UTC $\pm$ 1~h). The peak flux was $0.179^{+0.073}_{-0.057}$ meteoroids per 1000 km$^2$ per hour at a limiting magnitude of $+5.42$ and limiting mass of $1.5 \times 10^{-5}$~kg. Translated to zenithal hourly rate (ZHR), the peak activity was $0.65^{+0.24}_{-0.20}$ meteors per hour. The observed activity period spanned solar longitudes between $259.735^{\circ}$ and $260.318708^{\circ}$ (09:06 UTC to 22:53 UTC), with the threshold ZHR of around 0.1 meteors per hour. The network had excellent observational coverage throughout the activity period and had no coverage gaps. The minimum time-area product (TAP) per bin is set to 15,000~km$^2$~h and the minimum number of meteors per bin is set to 10, allowing for a temporal resolution of about 1 hour. The bin duration was limited by the number of meteors in each bin and not the TAP, showing the scarcity of observed meteors.

\begin{figure}[!h]
    \centering
    \centering
    \includegraphics[width=.48\textwidth]{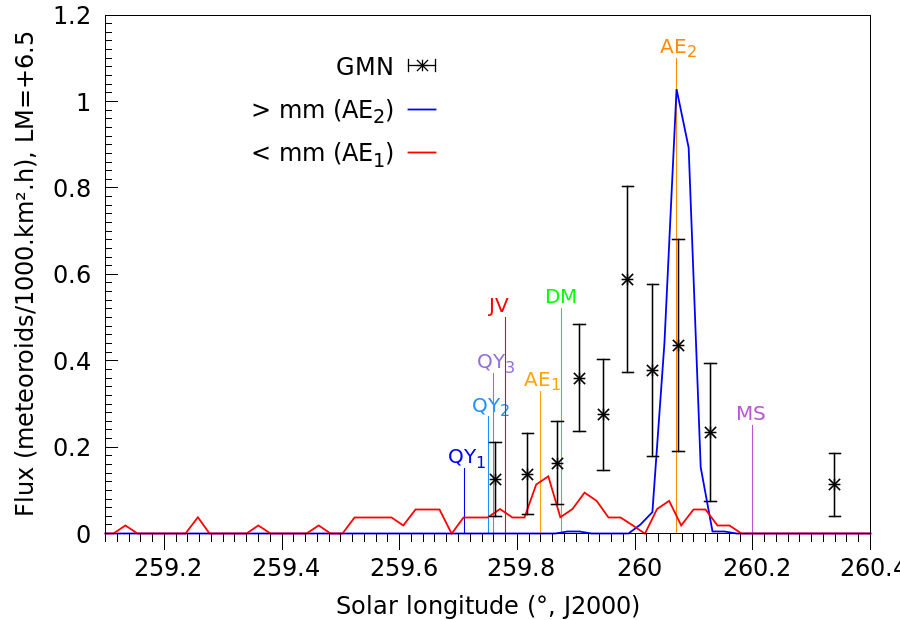}
    \caption{Comparison between the observed flux profile of the $\lambda$-Sculptorids with the predictions of \cite{vaubaillon2023new}. Coloured lines indicate the maximum activity time predicted by the different models presented in Table 2 of \cite{vaubaillon2023new}, and marked by the initials of the co-authors. The solid lines illustrate the modeled activity predicted by AE for particles with sizes above (blue) or below (red) one millimetre, with nodal-crossing locations closer than 0.02 AU from the Earth's orbit at the time of the shower (AE$_1$ model). The black symbols represent the flux measured by GMN at a limiting magnitude of +6.5.}
    \label{fig:predicted_profile}
\end{figure}

Figure \ref{fig:predicted_profile} shows the comparison between the predicted and observed activity profiles. \citet{vaubaillon2023new} found that different ejection conditions at the comet highly influence the shower's predicted time, which varied from 259.71$^{\circ}$ to 260.20$^{\circ}$ in solar longitude. In addition, the trails released by 46P display a complicated dynamical evolution that is typical of JFC meteoroid stream \citep{Vaubaillon2004}. This is particularly true for the 1974 trail, which suffered a close encounter with Jupiter in 1984 that significantly changed the meteoroids' orbital elements, causing the positions of the comet and the trail to reverse, putting the meteoroids in a leading position in front of the comet (see the comparison of the orbital elements before and after the encounter in Figure \ref{fig:ae2023}
).

All models in \cite{vaubaillon2023new} predicted a meteor activity caused by the 1974 trail. However, only the smallest meteoroids ($<$1 mm) of the trail were found to approach the Earth in 2023, with masses below the detection limit of GMN cameras. This discrepancy can be resolved if models considered lower particle densities than the predicted values of 500 to 1000 kg~m$^{-3}$. Additional simulations conducted by JV with a density of 100~kg~m$^{-3}$ show that massive particles from the 1974 trail are able to reach the Earth in 2023, with radiants and arrival times similar to those of \cite{vaubaillon2023new} and matching the observed radiant locations but not the peak time and activity profile. This low density is plausible as the shower's $K_B$ parameter is similar to that of the $\tau$-Herculids which also had a very low measured bulk density of 250~kg~m$^{-3}$ \citep{egal2023modeling}.

Additional investigations of the AE$_1$ simulations, motivated by the observational constraints, highlight the influence of the particles' selection criterion on the modeled activity profile. In the simulations, only particles crossing the ecliptic plane within a distance $\Delta X$ and time $\Delta T$ from the Earth are retained as potential impactors. For a standard $\Delta X < 0.02$~AU and $\Delta T\pm$ 7 days, only particles from the 1974 trail and the 1900-1908+ trails produced detectable meteors (model AE$_1$ in \cite{vaubaillon2023new}), as shown in Figure \ref{fig:predicted_profile}. 

\begin{figure}[!h]
    \centering
    \includegraphics[width=.48\textwidth]{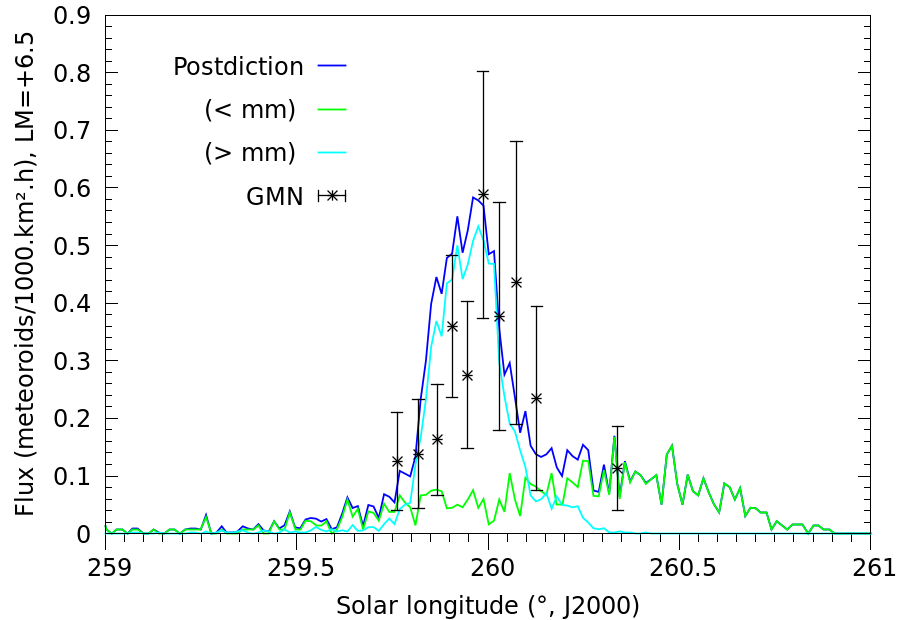}
    \caption{Updated modeled activity (postdiction) of the $\lambda$-Sculptorid shower from the AE$_1$ simulation data set, for sub-mm (green) and mm-sized (cyan) particles. In \cite{vaubaillon2023new}, only particles crossing the ecliptic plane closer than $\Delta r=$0.02 AU and within $\Delta T \pm$ 7 days from the Earth were retained in AE$_1$'s simulation. By increasing the time selection criterion to 20 days, more mm-sized particles ejected from 46P prior to 1908 approached the Earth in 2023. The resulting activity profile (dark blue) is a better match to the observations when including the contribution of older trails. }
    \label{fig:postdicted_profile}
\end{figure}

By increasing $\Delta T$ to 20 days, we find the modeled profile to be in better agreement with the observations (blue curve in Figure \ref{fig:postdicted_profile}). The temporal  $\Delta T$ selection amounts to replacing each simulated particle $p$ with a swarm of particles of identical orbits centred on $p$, which crosses the ecliptic plane at the nodal during a period of 2$\Delta T$. Large $\Delta T$ values, representing a high uncertainty on the particles' position along the orbits, have been employed to successfully predict the return of past meteor showers \citep{Jenniskens2008b,Egal2020}. 

With this less restrictive $\Delta T$ criterion, several trails ejected prior to 1908 are found to contribute to the shower's activity in 2023. The resulting activity profile reproduces well the timing and duration reported for the shower. In contrast, the modeled radiants remained similar to those presented in Figure \ref{fig:radiants_comparison}, and do not reproduce the observed spread in right ascension as well as the JV model. While additional observations could help assess the real contribution of trails ejected from the comet prior to 1974, no significant activity of the $\lambda$-Sculptorid is expected until 2045 in the AE$_1$ model.

\section{Conclusions}

We have presented the first observations of meteors from a new meteor shower $\lambda$-Sculptorids produced by the comet 46P/Wirtanen. The dynamical simulations by \citet{vaubaillon2023new} predicted that meteoroids from the 1974 trail encounter the Earth on December 12, 2023. The authors used several different methods and thus the exact predicted time of the peak varied between 08:23 UT and 20:06 UT.

A total of 23 multi-station meteors were observed by video meteor cameras of the Global Meteor Network for which trajectories were correlated, and a total of 172 from single-station $\lambda$-Sculptorids were used to estimate the flux. The activity peaked at 2023-12-12 15:04 UTC $\pm$ 1~h (solar longitude $\lambda_{\odot} = 259.988^{\circ} \pm 0.042^{\circ}$), while the total activity period (ZHR $>$ 0.1) spanned between 09:06 UTC to 22:53 UTC (solar longitudes $259.735^{\circ}$ and $260.318708^{\circ}$). The shower peaked at a ZHR of $0.65^{+0.24}_{-0.20}$ meteors per hour.

Due to the slow in-atmosphere speed of $\sim 15$~km~s$^{-1}$, only the largest meteoroids were observed, with an average size of 10~mm. The measured value of $K_B = 7.0 \pm 0.1$ indicates an ordinary short-period comet composition, similarly to the recent $\tau$-Herculid outburst in 2022, which was produced by similarly young meteoroids ejected in 1995. 

Trajectory, orbit, and activity level observations by the Global Meteor Network all matched the meteoroid stream dynamical simulations well overall. The radiants best match the predictions of JV for the 1974 trail if a very low bulk density of 100~kg~m$^{-3}$ is assumed. For higher densities, mm-sized and larger meteoroids do not reach the Earth in 2023 in all dynamical simulations. Alternatively, the observed activity profile and the mm-sizes can be achieved using older trails and if a larger threshold of time and range from the orbit is taken into account (i.e., AE model). However, this model reproduces the radiant structure less accurately than other models. The fundamental difficulty in reproducing the observations stems from the close encounter of the 1974 trail with Jupiter in 1984, which caused a significant shift in the orbital parameters. Any small model uncertainties are amplified by that event, setting a practical limit on the reproducibility of the observations. This makes the $\lambda$-Sculptorids an excellent data point for future model validation and improvement, as small differences in the model assumptions make large differences in the final result, helping to identify the best approach and model parameters.

\section*{Acknowledgements}\label{sec:ack}

The Global Meteor Network (GMN) data are released under the CC BY 4.0 license. The authors thank all GMN camera operators and contributors who are listed in Appendix \ref{app:gmn_members}. We also thank Peter Brown for providing CMOR observations of the new shower.

QY was supported by NASA grant 80NSSC21K0156, DV and AE were supported in part by the NASA Meteoroid Environment Office under cooperative agreement 80NSSC21M0073.

\bibliographystyle{aa} 
\bibliography{main}

\begin{appendix}

\clearpage
\onecolumn

\begin{landscape}
\section{Meteor trajectories and orbits}
\begin{longtable}{lrrrrrrrrrrrrrrrr}
\caption{Trajectory and orbit parameters of the observed meteors by the Global Meteor Network.}
\label{tab:trajectories} \\
\hline
Time (UTC) & $\lambda_{\odot}$ & $\alpha_{g}$ & $\delta_{g}$ & $V_{g}$ & $a$ & $e$ & $i$ & $q$ & $\Omega$ & $\omega$ & $\pi$ & $h_{B}$ & $h_{E}$ & $h_{\mathrm{peak}}$ & $M_{\mathrm{peak}}$ & $K_{B}$ \\
                       & (deg) & (deg) & (deg) & (km/s) & (au) & & (deg) & (au) & (deg) & (deg) & (deg) & (km) & (km) & (km) & (mag) & \\
\hline
\endfirsthead
\caption{continued.} \\
\hline
Time (UTC) & $\lambda_{\odot}$ & $\alpha_{g}$ & $\delta_{g}$ & $V_{g}$ & $a$ & $e$ & $i$ & $q$ & $\Omega$ & $\omega$ & $\pi$ & $h_{B}$ & $h_{E}$ & $h_{\mathrm{peak}}$ & $M_{\mathrm{peak}}$ & $K_{B}$ \\
                       & (deg) & (deg) & (deg) & (km/s) & (au) & & (deg) & (au) & (deg) & (deg) & (deg) & (km) & (km) & (km) & (mag) & \\
\hline
\endhead 
\hline
\endfoot
\hline
\multicolumn{17}{{p{\linewidth}}}{\footnotesize All meteors were observed on December 12. Standard deviations of individual parameters are given in alternating rows. $\lambda_{\odot}$ is the solar longitude, $\alpha_g$ and $\delta_g$ are geocentric right ascension and declination in J2000, $V_g$ is the geocentric velocity, $a$ is the semi-major axis, $e$ is the eccentricity, $i$ is the inclination, $q$ is the perihelion distance, $\Omega$ is the ascending node, $\omega$ is the argument of perihelion, $\pi$ is the longitude of perihelion, $h_B$ is the meteor being height, $h_E$ is the end height, $h_{\mathrm{peak}}$ is the height of peak magnitude, $M_{\mathrm{peak}}$ is the peak magnitude, and $K_{B}$ is the \citet{ceplecha1958composition} parameter.} \\
\endlastfoot
09:41:46.307 & 259.760586 &    7.84 &   -38.62 & 10.392 &  3.140 &  0.6864 &  9.439 & 0.9846310 &  79.768747 &    0.30 &   80.07 &  90.218 &  76.819 &  84.199 & +0.21 & 6.94 \\ 
             &            &    0.32 &     0.18 &  0.074 &  0.070 &  0.0069 &  0.038 & 0.0000000 &            &    0.14 &    0.14 &   0.050 &   0.040 &         &       &      \\ 
10:30:52.915 & 259.795268 &    7.16 &   -38.43 &  9.900 &  2.832 &  0.6523 &  8.991 & 0.9846280 &  79.802844 &    0.08 &   79.88 &  89.853 &  78.091 &  81.758 & +0.46 & 6.97 \\ 
             &            &    0.21 &     0.20 &  0.061 &  0.032 &  0.0040 &  0.079 & 0.0000000 &            &    0.10 &    0.10 &   0.030 &   0.020 &         &       &      \\ 
10:45:53.244 & 259.805864 &    7.76 &   -39.13 & 10.579 &  3.254 &  0.6974 &  9.668 & 0.9846250 &  79.813104 &    0.07 &   79.88 &  89.906 &  81.326 &  86.231 & +1.65 & 7.00 \\ 
             &            &    1.77 &     1.10 &  0.684 &  0.901 &  0.0542 &  0.447 & 0.0002000 &            &    1.08 &    1.08 &   0.160 &   0.170 &         &       &      \\ 
10:55:57.111 & 259.812971 &    5.82 &   -40.22 & 10.359 &  3.060 &  0.6782 &  9.558 & 0.9845520 &  79.819086 &  358.87 &   78.69 &  89.323 &  77.529 &  81.388 & -0.07 & 7.03 \\ 
             &            &    0.63 &     0.77 &  0.085 &  0.046 &  0.0049 &  0.187 & 0.0000000 &            &    0.44 &    0.44 &   0.060 &   0.060 &         &       &      \\ 
11:05:57.961 & 259.820043 &    8.29 &   -39.09 & 10.100 &  2.896 &  0.6601 &  9.365 & 0.9846180 &  79.826519 &    0.34 &   80.17 &  88.731 &  77.457 &  81.778 & -0.45 & 7.07 \\ 
             &            &    0.59 &     0.44 &  0.140 &  0.131 &  0.0130 &  0.065 & 0.0000000 &            &    0.33 &    0.33 &   0.050 &   0.030 &         &       &      \\ 
12:00:04.960 & 259.858260 &    6.56 &   -38.65 &  9.861 &  2.808 &  0.6494 &  8.955 & 0.9846150 &  79.864074 &  359.71 &   79.58 &  91.385 &  77.007 &  87.773 & +1.77 & 6.86 \\ 
             &            &    0.42 &     0.14 &  0.134 &  0.067 &  0.0084 &  0.152 & 0.0000000 &            &    0.14 &    0.14 &   0.070 &   0.060 &         &       &      \\ 
12:46:23.484 & 259.890963 &    6.58 &   -39.45 &  9.782 &  2.722 &  0.6383 &  9.057 & 0.9846060 &  79.902111 &  359.45 &   79.35 &  92.840 &  77.232 &  83.906 & +0.20 & 6.75 \\ 
             &            &    0.19 &     0.33 &  0.095 &  0.070 &  0.0088 &  0.021 & 0.0000000 &            &    0.18 &    0.18 &   0.070 &   0.030 &         &       &      \\ 
12:57:12.681 & 259.898605 &    6.79 &   -39.27 &  9.831 &  2.755 &  0.6426 &  9.076 & 0.9846140 &  79.909016 &  359.59 &   79.50 &  90.072 &  72.864 &  81.493 & -0.44 & 6.94 \\ 
             &            &    0.33 &     0.17 &  0.057 &  0.043 &  0.0058 &  0.016 & 0.0000000 &            &    0.13 &    0.13 &   0.020 &   0.020 &         &       &      \\ 
13:01:31.386 & 259.901649 &    6.27 &   -37.72 &  9.848 &  2.853 &  0.6549 &  8.739 & 0.9846130 &  79.907307 &  359.85 &   79.76 &  89.511 &  81.264 &  85.819 & +0.95 & 7.03 \\ 
             &            &    0.56 &     0.99 &  0.158 &  0.052 &  0.0061 &  0.336 & 0.0000000 &            &    0.24 &    0.24 &   0.070 &   0.060 &         &       &      \\ 
13:07:48.998 & 259.906093 &    6.26 &   -37.86 &  9.996 &  2.942 &  0.6653 &  8.868 & 0.9846170 &  79.916954 &  359.78 &   79.70 &  89.483 &  77.391 &  83.160 & +1.26 & 7.00 \\ 
             &            &    0.27 &     1.65 &  0.440 &  0.344 &  0.0435 &  0.033 & 0.0000000 &            &    0.42 &    0.42 &   0.130 &   0.130 &         &       &      \\ 
13:14:46.735 & 259.911011 &    6.89 &   -39.91 &  9.730 &  2.666 &  0.6307 &  9.134 & 0.9846020 &  79.920482 &  359.43 &   79.35 &  88.251 &  79.740 &  83.293 & +1.11 & 7.06 \\ 
             &            &    0.24 &     0.41 &  0.097 &  0.033 &  0.0048 &  0.154 & 0.0000000 &            &    0.14 &    0.14 &   0.020 &   0.010 &         &       &      \\ 
13:21:43.108 & 259.915911 &    7.09 &   -39.44 &  9.903 &  2.783 &  0.6462 &  9.186 & 0.9846120 &  79.925115 &  359.66 &   79.58 &  92.288 &  74.862 &  81.079 & -0.43 & 6.78 \\ 
             &            &    0.19 &     0.18 &  0.024 &  0.003 &  0.0003 &  0.066 & 0.0000000 &            &    0.06 &    0.06 &   0.020 &   0.020 &         &       &      \\ 
13:22:48.440 & 259.916680 &    6.04 &   -38.38 &  9.692 &  2.732 &  0.6396 &  8.740 & 0.9846030 &  79.921885 &  359.55 &   79.47 &  91.239 &  80.946 &  85.689 & +0.92 & 6.89 \\ 
             &            &    0.25 &     0.15 &  0.015 &  0.012 &  0.0016 &  0.045 & 0.0000000 &            &    0.11 &    0.11 &   0.030 &   0.030 &         &       &      \\ 
13:30:12.540 & 259.921907 &   10.80 &   -38.57 & 10.269 &  2.972 &  0.6688 &  9.610 & 0.9844630 &  79.926941 &    1.57 &   81.49 &  89.407 &  82.104 &  84.754 & +2.40 & 7.06 \\ 
             &            &    2.30 &     0.28 &  0.182 &  0.060 &  0.0074 &  0.366 & 0.0001000 &            &    0.95 &    0.95 &   0.150 &   0.120 &         &       &      \\ 
13:44:07.285 & 259.931732 &    9.26 &   -37.94 & 10.364 &  3.122 &  0.6847 &  9.411 & 0.9845450 &  79.940239 &    1.07 &   81.01 &  90.921 &  78.231 &  84.105 & +0.49 & 6.90 \\ 
             &            &    0.65 &     1.74 &  0.466 &  0.186 &  0.0245 &  0.697 & 0.0001000 &            &    0.32 &    0.32 &   0.190 &   0.060 &         &       &      \\ 
14:39:26.951 & 259.970805 &    6.97 &   -39.04 & 10.235 &  3.020 &  0.6739 &  9.335 & 0.9846000 &  79.978607 &  359.68 &   79.66 &  90.153 &  79.277 &  86.483 & +0.80 & 6.97 \\ 
             &            &    0.20 &     0.15 &  0.034 &  0.011 &  0.0012 &  0.068 & 0.0000000 &            &    0.07 &    0.07 &   0.020 &   0.010 &         &       &      \\ 
15:02:38.789 & 259.987187 &    6.28 &   -39.95 &  9.931 &  2.789 &  0.6470 &  9.236 & 0.9846210 &  80.002258 &  359.10 &   79.10 &  90.100 &  71.945 &  76.310 & -0.26 & 7.17 \\ 
             &            &    0.12 &     0.09 &  0.026 &  0.018 &  0.0023 &  0.011 & 0.0000000 &            &    0.08 &    0.08 &   0.070 &   0.040 &         &       &      \\ 
15:24:13.360 & 260.002425 &    6.26 &   -37.78 & 10.088 &  3.009 &  0.6728 &  8.915 & 0.9845980 &  80.009246 &  359.75 &   79.76 &  88.584 &  80.556 &  85.417 & +0.99 & 7.09 \\ 
             &            &    0.72 &     0.75 &  0.072 &  0.016 &  0.0017 &  0.258 & 0.0000000 &            &    0.18 &    0.18 &   0.040 &   0.050 &         &       &      \\ 
15:32:00.210 & 260.007921 &    9.66 &   -38.34 &  9.938 &  2.804 &  0.6489 &  9.221 & 0.9845250 &  80.015119 &    1.11 &   81.12 &  88.757 &  78.596 &  83.334 & +0.88 & 7.07 \\ 
             &            &    1.06 &     0.29 &  0.026 &  0.010 &  0.0013 &  0.064 & 0.0000000 &            &    0.55 &    0.55 &   0.040 &   0.070 &         &       &      \\ 
16:06:14.321 & 260.032097 &    7.38 &   -39.37 & 10.243 &  2.996 &  0.6714 &  9.442 & 0.9846270 &  80.038368 &  359.73 &   79.77 &  91.896 &  83.424 &  88.004 & +1.24 & 6.96 \\ 
             &            &    0.40 &     0.30 &  0.121 &  0.077 &  0.0085 &  0.109 & 0.0000000 &            &    0.27 &    0.27 &   0.050 &   0.040 &         &       &      \\ 
16:33:46.541 & 260.051545 &    6.75 &   -39.84 & 10.218 &  2.965 &  0.6679 &  9.464 & 0.9845640 &  80.057597 &  359.29 &   79.35 &  90.695 &  76.727 &  85.900 & +0.13 & 6.94 \\ 
             &            &    0.67 &     0.08 &  0.047 &  0.041 &  0.0049 &  0.022 & 0.0000000 &            &    0.32 &    0.32 &   0.040 &   0.080 &         &       &      \\ 
16:55:42.027 & 260.067029 &    8.74 &   -38.89 & 10.120 &  2.909 &  0.6616 &  9.381 & 0.9845780 &  80.073024 &    0.47 &   80.55 &  90.502 &  81.624 &  84.233 & +1.90 & 6.95 \\ 
             &            &    1.00 &     0.20 &  0.031 &  0.011 &  0.0012 &  0.080 & 0.0000000 &            &    0.50 &    0.50 &   0.070 &   0.060 &         &       &      \\ 
17:11:49.569 & 260.078417 &    6.08 &   -38.88 &  9.610 &  2.663 &  0.6302 &  8.780 & 0.9845680 &  80.084380 &  359.33 &   79.42 &  90.613 &  82.306 &  85.386 & +1.35 & 6.92 \\ 
             &            &    0.72 &     0.28 &  0.066 &  0.033 &  0.0044 &  0.071 & 0.0000000 &            &    0.39 &    0.39 &   0.080 &   0.010 &         &       &      \\
\end{longtable}

\begin{longtable}{rrrrrrrrrr}
\caption{Flux measurements by the GMN.}
\label{tab:flux} \\
\hline
$\lambda_{\odot}$ & Flux(+6.5) &  Flux(+5.42) &  ZHR &  Meteor Count &  TAP &  Meteor LM &  Radiant elev. &  Radiant dist. &  Ang vel. \\
(deg) & (met / 1000 km$^2$ h) & (met / 1000 km$^2$ h) & (met/h) & & (1000 km$^2$ h) & (mag) & (deg) & (deg) & (deg/s) \\
\hline
\endfirsthead
\caption{continued.} \\
\hline
$\lambda_{\odot}$ & Flux@+6.5M &  Flux@+5.42M &  ZHR &  Meteor Count &  TAP &  Meteor LM &  Radiant elev. &  Radiat dist. &  Ang vel. \\
(deg) & (met / 1000 km$^2$ h) & (met / 1000 km$^2$ h) & (met/h) & & (1000 km$^2$ h) & (mag) & (deg) & (deg) & (deg/s) \\
\hline
\endhead 
\hline
\endfoot
\hline
\multicolumn{10}{p{\linewidth}}{\footnotesize $\lambda_{\odot}$ is the time-area product (TAP) weighted solar longitude of the bin. The two flux columns are fluxes at the standard magnitude of +6.5 and the measured mean magnitude of +5.42. Meteor LM is the mean TAP-weighted apparent meteor limiting magnitude, followed by the mean radiant parameters weighted in the same way.} \\
\endlastfoot
259.7351911 & $0.125_{-0.065}^{+0.106}$ & $0.038_{-0.020}^{+0.032}$ & $0.138_{-0.072}^{+0.115}$ & 10 & 79.760 & 5.36 & 52.58 & 61.34 & 6.08 \\
259.8190395 & $0.138_{-0.072}^{+0.117}$ & $0.042_{-0.022}^{+0.035}$ & $0.152_{-0.079}^{+0.127}$ & 10 & 72.255 & 5.45 & 43.28 & 64.32 & 6.37 \\
259.8635306 & $0.163_{-0.076}^{+0.116}$ & $0.050_{-0.024}^{+0.035}$ & $0.179_{-0.084}^{+0.127}$ & 13 & 79.795 & 5.46 & 53.29 & 60.99 & 6.03 \\
259.9063104 & $0.360_{-0.108}^{+0.139}$ & $0.110_{-0.033}^{+0.042}$ & $0.395_{-0.118}^{+0.152}$ & 36 & 99.887 & 5.53 & 56.61 & 58.49 & 5.87 \\
259.9473790 & $0.275_{-0.107}^{+0.149}$ & $0.084_{-0.033}^{+0.045}$ & $0.301_{-0.117}^{+0.164}$ & 20 & 72.807 & 5.46 & 52.77 & 61.90 & 5.95 \\
259.9884477 & $0.588_{-0.186}^{+0.243}$ & $0.179_{-0.057}^{+0.074}$ & $0.646_{-0.204}^{+0.265}$ & 32 & 54.381 & 5.41 & 46.60 & 65.48 & 6.10 \\
260.0312275 & $0.377_{-0.162}^{+0.235}$ & $0.115_{-0.049}^{+0.071}$ & $0.413_{-0.177}^{+0.258}$ & 16 & 42.488 & 5.41 & 43.32 & 66.76 & 6.10 \\
260.0740073 & $0.435_{-0.197}^{+0.295}$ & $0.132_{-0.060}^{+0.090}$ & $0.477_{-0.216}^{+0.324}$ & 14 & 32.179 & 5.38 & 35.02 & 72.36 & 6.24 \\
260.1321879 & $0.234_{-0.122}^{+0.197}$ & $0.071_{-0.037}^{+0.060}$ & $0.257_{-0.134}^{+0.216}$ & 10 & 42.696 & 5.33 & 26.98 & 78.32 & 6.23 \\
260.3187080 & $0.113_{-0.056}^{+0.090}$ & $0.034_{-0.017}^{+0.028}$ & $0.124_{-0.062}^{+0.098}$ & 11 & 97.173 & 5.31 & 34.52 & 71.62 & 6.71 \\
\end{longtable}
\end{landscape}

\section{Additional figures}

\begin{figure}[H]
    \centering\includegraphics[width=0.8\linewidth]{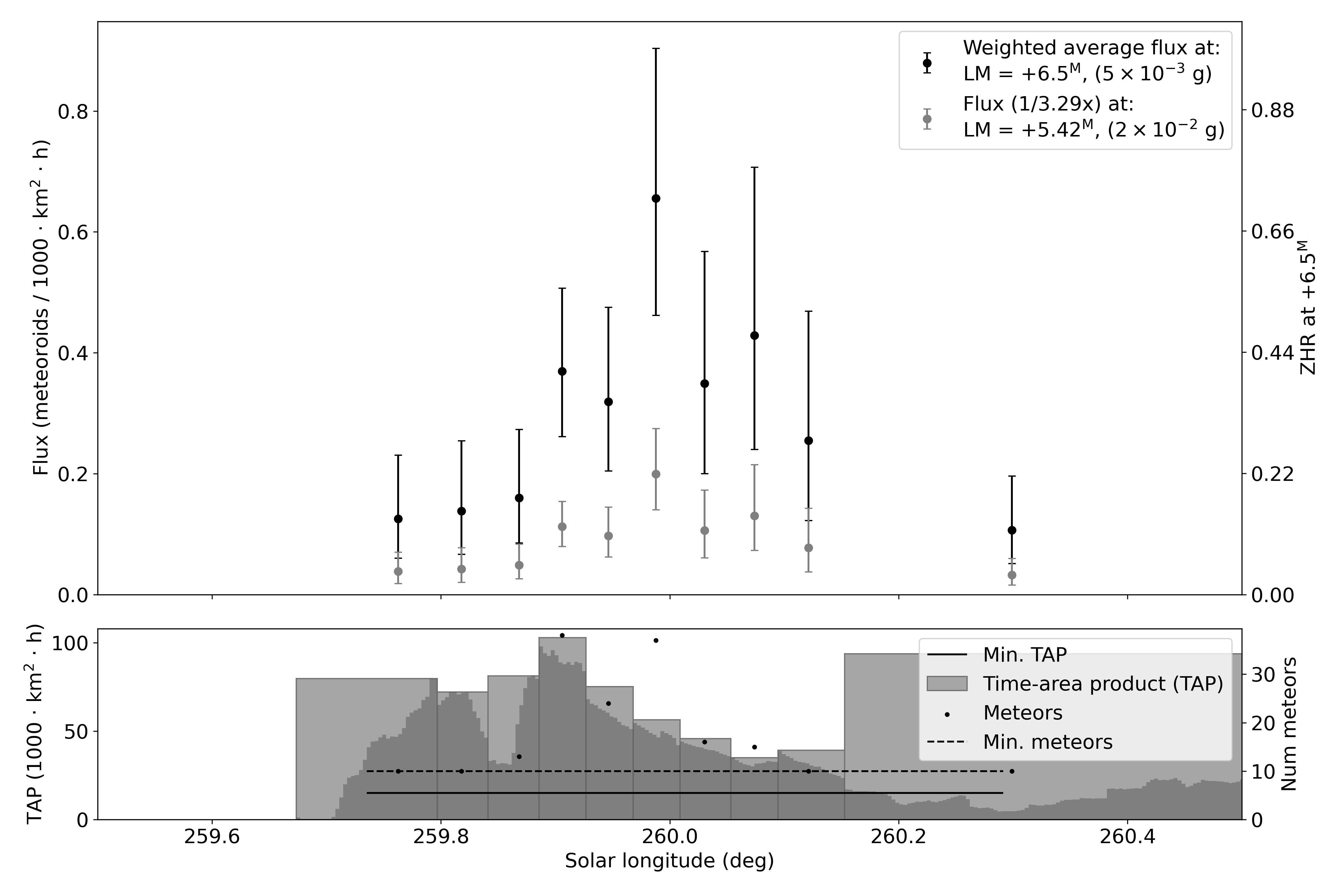}
    \caption{Flux profile of the $\lambda$-Sculpturids. The top inset shows the flux at two different limiting magnitudes: black markers for the standard $+6.5$ and gray markers for the observed mean of $+5.42$. The bottom inset shows the observed time-area product (TAP) and the observed number of meteors, together with the thresholds set on both. The dark gray shading inside individual TAP bins shows the distribution of TAP inside each bin, revealing any potential gaps in coverage.}
    \label{fig:gmn_flux}
\end{figure}

\begin{figure}[!h]
    \centering
    \includegraphics[width=0.7\textwidth]{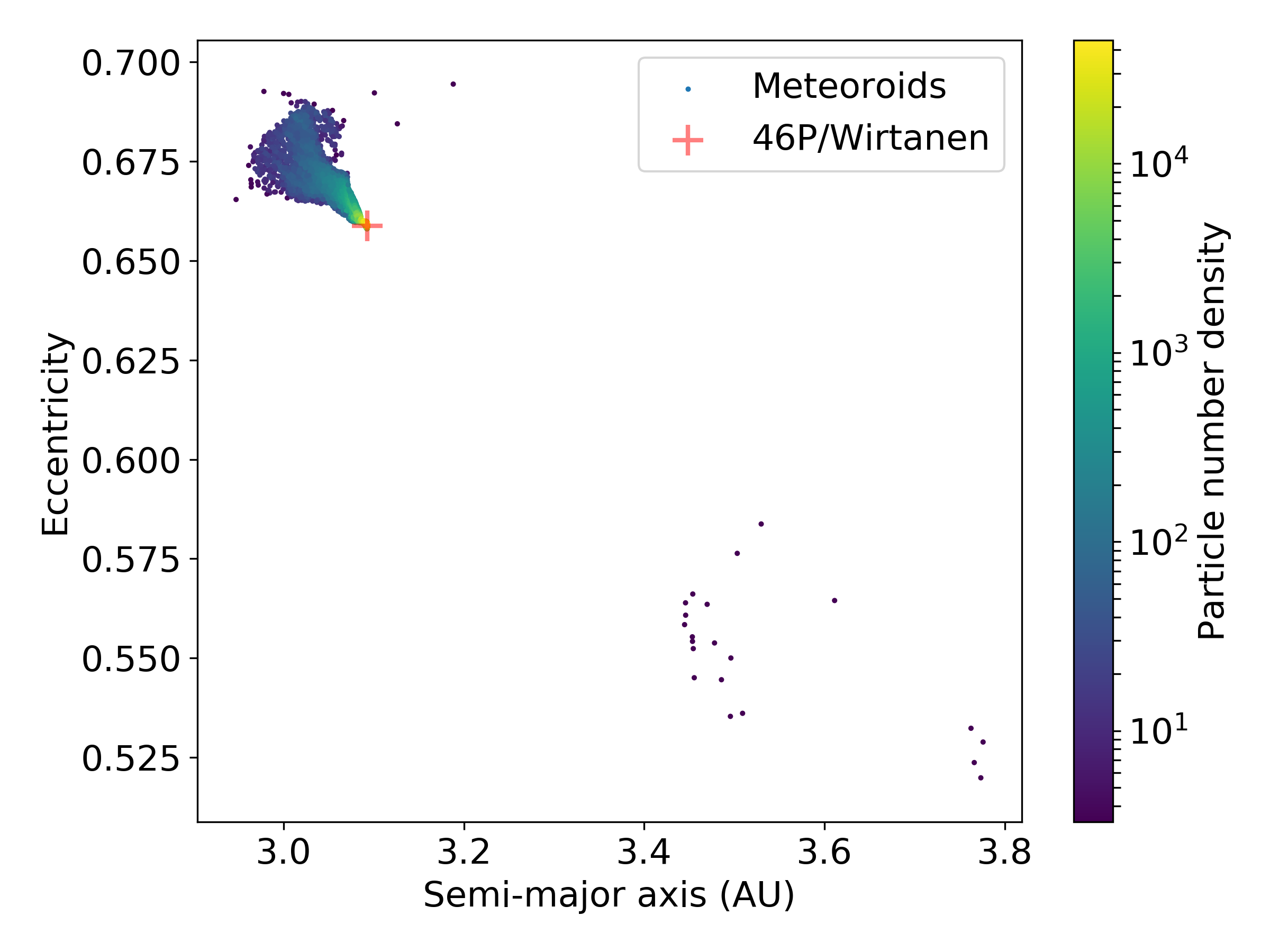}
    \caption{Orbits of simulated particles encountering the Earth in 2023, ejected in 1974 (JV model). Particles are coloured by the relative number density. The red marker is the orbit of 46P/Wirtanen. The close encounter with Jupiter in 1984 caused the so-called reversal process, eventually making the meteoroids lead the parent.}
    \label{fig:ae2023}
\end{figure}

\clearpage
\twocolumn

\section{Extended acknowledgements} \label{app:gmn_members}

The authors thank all Global Meteor Network camera operators and contributors:

\noindent Adam Mullins, Aden Walker, Adrian Bigland, Adriana Roggemans, Alain Marin, Alan Beech, Alan Maunder, Alan Pevec, Alan Pickwick, Alan Decamps, Alan Cowie, Aled Powell, Alejandro Barriuso, Aleksandar Merlak, Alex Bell, Alex Haislip, Alex Hodge, Alex Jeffery, Alex Kichev, Alex McConahay, Alex Pratt, Alex Roig, Alex Aitov, Alexander Wiedekind-Klein, Alexandre Alves, Alfredo Dal' Ava Júnior, Amy Barron, Anatoly Ijon, Andre Rousseau, Andre Bruton, Andrea Storani, Andrei Marukhno, Andres Fernandez, Andrew Campbell-Laing, Andrew Challis, Andrew Cooper, Andrew Fiamingo, Andrew Heath, Andrew Moyle, Andrew Washington, Andy Stott, Andy Sapir, Ange Fox, Angel Sierra, Angélica López Olmos, Ansgar Schmidt, Anthony Hopkinson, Anthony Pitt, Anthony Kesterton, Anton Macan, Anton Yanishevskiy, Anzhari Purnomo, Arie Blumenzweig, Arie Verveer, Arnaud Leroy, Attila Nemes, Barry Findley, Bart Dessoy, Bela Szomi Kralj, Bernard Côté, Bernard Hagen, Bev M. Ewen-Smith, Bill Cooke, Bill Wallace, Bill Witte, Bill Carr, Bob Evans, Bob Greschke, Bob Hufnagel, Bob Marshall, Bob Massey, Bob Zarnke, Bob Guzik, Brenda Goodwill, Brendan Cooney, Brian Chapman, Brian Murphy, Brian Rowe, Brian Hochgurtel, Wyatt Hochgurtel, Brian Mitchell, Bruno Bonicontro, Callum Potter, Carl Elkins, Carl Mustoe, Carl Panter, Charles Thody, Charlie McCormack, Chris Baddiley, Chris Blake, Chris Dakin, Chris George, Chris James, Chris Ramsay, Chris Reichelt, Chris Chad, Chris O'Neill, Chris White, Christian Wanlin, Christine Ord, Christof Zink, Christophe Demeautis, Christopher Coomber, Christopher Curtis, Christopher Tofts, Christopher Brooks, Chuck Goldsmith, Chuck Pullen, Ciaran Tangney, Claude Boivin, Claude Surprenant, Clive Sanders, Colin Graham, Colin Marshall, Colin Nichols, Con Stoitsis, Creina Beaman, Daknam Al-Ahmadi, Damien Lemay, Damien McNamara, Damir Matković, Damir Šegon, Damjan Nemarnik, Dan Klinglesmith, Dan Pye, Daniel Duarte, Daniel J. Grinkevich, Daniela Cardozo Mourão, Danijel Reponj, Danko Kočiš, Dario Zubović, Dave Jones, Dave Mowbray, Dave Newbury, Dave Smith, David Akerman, David Attreed, David Bailey, David Brash, David Castledine, David Hatton, David Leurquin, David Price, David Rankin, David Robinson, David Rollinson, David Strawford, David Taylor, David Rogers, David Banes, David Johnston, David Rees, Dean Moore, Denis Bergeron, Denis St-Gelais, Dennis Behan, Derek Poulton, Didier Walliang, Dimitris Georgoulas, Dino Čaljkušić, Dmitrii Rychkov, Dominique Guiot, Don Anderson, Don Hladiuk, Dorian Božičević, Dougal Matthews, Douglas Sloane, Douglas Stone, Dustin Rego, Dylan O’Donnell, Ed Breuer, Ed Harman, Edd Stone, Edgar Mendes Merizio, Edison José Felipe Pérezgómez Álvarez, Edson Valencia Morales, Eduardo Fernandez Del Peloso, Edward Cooper, Ehud Behar, Eleanor Mayers, Enrico Pettarin, Enrique Arce, Enrique Chávez Garcilazo, Eric Lopez, Eric Toops, Erwin van Ballegoij, Erwin Harkink, Eugene Potapov, Ewan Richardson, Fabricio Borges, Fernando Dall'Igna, Fernando Jordan, Fernando Requena, Filip Matković, Filip Mezak, Filip Parag, Fiona Cole, Florent Benoit, Francis Rowsell, François Simard, Frank Lyter, Frantisek Bilek, Gabor Sule, Gaétan Laflamme, Gareth Brown, Gareth Lloyd, Gareth Oakey, Garry Dymond, Gary Parker, Gary Eason, Gavin Martin, Gene Mroz, Geoff Scott, Georges Attard, Georgi Momchilov, Germano Soru, Gilton Cavallini, Gordon Hudson, Graeme Hanigan, Graham Stevens, Graham Winstanley, Graham Henstridge, Greg Michael, Gustav Frisholm, Gustavo Silveira B. Carvalho, Guy Létourneau, Guy Williamson, Guy Lesser, Hamish Barker, Hamish McKinnon, Haris Jeffrey, Harri Kiiskinen, Hartmut Leiting, Heather Petelo, Heriton Rocha, Hervé Lamy, Herve Roche, Holger Pedersen, Horst Meyerdierks, Howard Edin, Hugo González, Iain Drea, Ian Enting Graham, Ian Lauwerys, Ian Parker, Ian Pass, Ian A. Smith, Ian Williams, Ian Hepworth, Ian Collins, Igor Duchaj, Igor Henrique, Igor Macuka, Igor Pavletić, Ilya Jankowsky, Ioannis Kedros, Ivan Gašparić, Ivan Sardelić, Ivica Ćiković, Ivica Skokić, Ivo Dijan, Ivo Silvestri, Jack Barrett, Jacques Masson, Jacques Walliang, Jacqui Thompson, James Davenport, James Farrar, James Scott, James Stanley, James Dawson, Jamie Allen, Jamie Cooper, Jamie McCulloch, Jamie Olver, Jamie Shepherd, Jan Hykel, Jan Wisniewski, Janis Russell, Janusz Powazki, Jason Burns, Jason Charles, Jason Gill, Jason van Hattum, Jason Sanders, Javor Kac, Jay Shaffer, Jean Francois Larouche, Jean Vallieres, Jean Brunet, Jean-Baptiste Kikwaya, Jean-Louis Naudin, Jean-Marie Jacquart, Jean-Paul Dumoulin, Jean-Philippe Barrilliot, Jeff Holmes, Jeff Huddle, Jeff Wood, Jeff Devries, Jeffrey Legg, Jeremy Taylor, Jesse Stayte, Jesse Lard, Jessica Richards, Jim Blackhurst, Jim Cheetham, Jim Critchley, Jim Fordice, Jim Gilbert, Jim Rowe, Jim Seargeant, Jochen Vollsted, Jocimar Justino, John W. Briggs, John Drummond, John Hale, John Kmetz, John Maclean, John Savage, John Thurmond, John Tuckett, John Waller, John Wildridge, John Bailey, Jon Bursey, Jonathan Alexis Valdez Aguilar, Jonathan Eames, Jonathan Mackey, Jonathan Whiting, Jonathan Wyatt, Jonathon Kambulow, Jorge Augusto Acosta Bermúdez, Jorge Oliveira, Jose Carballada, Jose Galindo Lopez, José María García, José-Luis Martín, Josip Belas, Josip Krpan, Jost Jahn, Juan Luis Muñoz, Jure Zakrajšek, Jürgen Dörr, Jürgen Ketterer, Justin Zani, Karen Smith, Kath Johnston, Kees Habraken, Keith Maslin, Keith Biggin, Ken Jamrogowicz, Ken Lawson, Ken Gledhill, Kevin Gibbs-Wragge, Kevin Morgan, Klaas Jobse, Korado Korlević, Kyle Francis, Lachlan Gilbert, Larry Groom, Laurent Brunetto, Laurie Stanton, Lawrence Saville, Lee Hill, Len North, Leslie Kaye, Lev Pustil’Nik, Lexie Wallace, Lisa Holstein, Llewellyn Cupido, Lorna McCalman, Louw Ferreira, Lovro Pavletić, Lubomir Moravek, Luc Turbide, Lucia Dowling, Luciano Miguel Diniz, Ludger Börgerding, Luis Fabiano Fetter, Maciej Reszelsk, Magda Wisniewska, Manel Colldecarrera, Marc Corretgé Gilart, Marcel Berger, Marcelo Domingues, Marcelo Zurita, Marcio Malacarne, Marco Verstraaten, Margareta Gumilar, Marián Harnádek, Mariusz Adamczyk, Mark Fairfax, Mark Gatehouse, Mark Haworth, Mark McIntyre, Mark Phillips, Mark Robbins, Mark Spink, Mark Suhovecky, Mark Williams, Mark Ward, Marko Šegon, Marthinus Roos, Martin Breukers, Martin Richmond-Hardy, Martin Robinson, Martin Walker, Martin Woodward, Martin Connors, Martyn Andrews, Mason McCormack, Mat Allan, Matej Mihelčić, Matt Cheselka, Matthew Howarth, Megan Gialluca, Mia Boothroyd, Michael Cook, Michael Mazur, Michael O’Connell, Michael Krocil, Michał Warchoł, Michel Saint-Laurent, Miguel Diaz Angel, Miguel Preciado, Mike Breimann, Mike Hutchings, Mike Read, Mike Shaw, Mike Ball, Milan Kalina, Minesh Patel, Mirjana Malarić, Muhammad Luqmanul Hakim Muharam, Murray Forbes, Murray Singleton, Murray Thompson, Myron Valenta, Nawaz Mahomed, Ned Smith, Nedeljko Mandić, Neil Graham, Neil Papworth, Neil Waters, Neil Petersen, Nelson Moreira, Neville Vann, Nial Bruce, Nicholas Hill, Nicholas Ruffier, Nick Howarth, Nick James, Nick Moskovitz, Nick Norman, Nick Primavesi, Nick Quinn, Nick Russel, Nicola Masseroni, Nigel Bubb, Nigel Evans, Nigel Owen, Nigel Harris, Nikola Gotovac, Nikolay Gusev, Nikos Sioulas, Noah Simmonds, Ollie Eisman, Pablo Canedo, Paraksh Vankawala, Pat Devine, Patrick Franks, Patrick Poitevin, Patrick Geoffroy, Patrik Kukić, Paul Cox, Paul Dickinson, Paul Haworth, Paul Heelis, Paul Kavanagh, Paul Ludick, Paul Prouse, Paul Pugh, Paul Roche, Paul Roggemans, Paul Stewart, Paul Huges, Pedro Augusto Hay Day, Penko Yordanov, Pete Graham, Pete Lynch, Peter G. Brown, Peter Campbell-Burns, Peter Davis, Peter Eschman, Peter Gural, Peter Hallett, Peter Jaquiery, Peter Kent, Peter Lee, Peter McKellar, Peter Meadows, Peter Stewart, Peter Triffitt, Peter Leigh, Pető Zsolt, Phil James, Philip Gladstone, Philip Norton, Philippe Schaak, Phillip Wilhelm Maximilian Grammerstorf, Pierre Gamache, Pierre de Ponthière, Pierre-Michael Micaletti, Pierre-Yves Pechart, Pieter Dijkema, Predrag Vukovic, Przemek Nagański, Radim Stano, Rajko Sušanj, Raju Aryal, Raoul van Eijndhoven, Raul Truta, Reinhard Kühn, Remi Lacasse, Renato Cássio Poltronieri, René Tardif, Richard Abraham, Richard Bassom, Richard Croy, Richard Davis, Richard Fleet, Richard Hayler, Richard Johnston, Richard Kacerek, Richard Payne, Richard Stevenson, Richard Severn, Rick Fischer, Rick Hewett, Rick James, Ricky Bassom, Rob Agar, Rob de Corday Long, Rob Saunders, Rob Smeenk, Robert Longbottom, Robert McCoy, Robert Saint-Jean, Robert D. Steele, Robert Veronneau, Robert Peledie, Robin Boivin, Robin Earl, Robin Rowe, Roel Gloudemans, Roger Banks, Roger Morin, Roland Idaczyk, Rolf Carstens, Romulo Jose, Ron James Jr, Roslina Hussain, Russell Jackson, Ryan Frazer, Ryan Harper, Salvador Aguirre, Sam Green, Sam Hemmelgarn, Sarah Tonorio, Scott Kaufmann, Sebastian Klier, Seppe Canonaco, Seraphin Feller, Serge Bergeron, Sergio Mazzi, Sevo Nikolov, Simon Cooke-Willis, Simon Holbeche, Simon Maidment, Simon McMillan, Simon Minnican, Simon Parsons, Simon Saunders, Simon Fidler, Sofia Ulrich, Srivishal Sudharsan, Stacey Downton, Stan Nelson, Stanislav Korotkiy, Stanislav Tkachenko, Stef Vancampenhout, Stefan Frei, Stephane Zanoni, Stephen Grimes, Stephen Nattrass, Steve Berry, Steve Bosley, Steve Carter, Steve Dearden, Steve Homer, Steve Kaufman, Steve Lamb, Steve Rau, Steve Tonkin, Steve Trone, Steve Welch, Steven Shanks, Steven Tilley, Stewart Doyle, Stuart Brett, Stuart Land, Stuart McAndrew, Sylvain Cadieux, Tammo Jan Dijkema, Terry Pundiak, Terry Richardson, Terry Simmich, Terry Young, Thiago Paes, Thomas Blog, Thomas Schmiereck, Thomas Stevenson, Tihomir Jakopčić, Tim Burgess, Tim Claydon, Tim Cooper, Tim Gloudemans, Tim Havens, Tim Polfliet, Tim Frye, Tioga Gulon, Tobias Westphal, Tom Warner, Tommy McEwan, Tommy B. Nielsen, Torcuill Torrance, Tosh White, Tracey Snelus, Trevor Clifton, Ubiratan Borges, Urs Wirthmueller, Uwe Glässner, Vasilii Savtchenko, Ventsislav Bodakov, Victor Acciari, Viktor Toth, Vincent McDermott, Vladimir Jovanović, Waily Harim, Warley Souza, Washington Oliveira, Wenceslao Trujillo, William Perkin, William Schauff, William Stewart, William Harvey, William Hernandez, Wullie Mitchell, Yakov Tchenak, Yfore Scott, Yohsuke Akamatsu, Yong-Ik Byun, Yozhi Nasvadi, Yuri Stepanychev, Zach Steele, Zané Smit, Zbigniew Krzeminski, Željko Andreić, Zhuoyang Chen, Zoran Dragić, Zoran Knez, Zoran Novak, Asociación de Astronomía de Marina Alta, Costa Blanca Astronomical Society, Perth Observatory Volunteer Group, Royal Astronomical Society of Canada Calgary Centre.

\end{appendix}

\end{document}